\documentclass[%
 reprint,
aps,
pra,
floatfix,
]{revtex4-1}

\usepackage{graphicx}
\usepackage{dcolumn}
\usepackage{bm}

\begin{document}


\title{Invertible mappings and the large deviation theory for the $q$-maximum entropy principle}

\author{R. C. Venkatesan}

 \email{ravi@systemsresearchcorp.com}
\affiliation{%
 Systems Research Corporation \\ITI Rd., Aundh, Pune 411007, India.\\
 }%

\author{A. Plastino}
\email{plastino@fisica.unlp.edu.ar}
\affiliation{
 IFLP, National University La Plata \&
National Research Council (CONICET) -  C. C., 727 1900, La Plata,
Argentina.\\
 }%

\date{\today}

\begin{abstract}
The possibility of reconciliation between canonical
probability distributions obtained from the $q$-maximum
entropy principle with predictions from the law of large numbers
when empirical samples are held to the same constraints, is investigated into. Canonical probability distributions are
constrained by both: $(i)$ the additive duality of generalized
statistics and $(ii)$  normal averages expectations.
Necessary conditions to establish such a reconciliation are derived by
appealing to a result concerning large deviation properties of
conditional measures. The (dual) $q^*$-maximum entropy principle
is shown {\bf not} to adhere to the large deviation theory.
However, the necessary conditions are proven to constitute an
invertible mapping between:  $(i)$ a canonical ensemble satisfying
the $q^*$-maximum entropy principle for energy-eigenvalues
$\varepsilon_i^*$, and, $(ii)$ a canonical ensemble satisfying the
Shannon-Jaynes maximum entropy theory for energy-eigenvalues
$\varepsilon_i$. Such an invertible mapping is demonstrated to
facilitate an \emph{implicit} reconciliation between the
$q^*$-maximum entropy principle and the large deviation theory.
Numerical examples for exemplary cases are provided.

\begin{description}
\item[PACS numbers]{05.20.-y, 02.50.Cw, 05.20.Gg}
\end{description}
\end{abstract}

\maketitle


\section{\label{sec:level1}Introduction}

The generalized (also, interchangeably, nonadditive or deformed)
statistics of Tsallis' has recently been the focus of much
attention in complex systems, and allied disciplines \cite{1}. The
generalized statistics of Tsallis' is adequate for statistical mechanical
systems exhibiting strong correlations and/or long-range
interactions. It has generated intense interest in physics and
allied disciplines. Many of the issues concerning the properties
q-statistics are a subject of intense debate (see, for example
\cite{2,3,4,5,6,7,8}), with questions, responses, and counter-responses by
many authors. A continually updated bibliography of
works related to the generalized statistics of Tsallis may be
found at http://tsallis.cat.cbpf.br/biblio.htm.

The large deviation theory constitutes an important statistical
basis for information entropies \cite{9,10}.  Following the pattern
concerning the properties and applicability of the generalized
statistics of Tsallis, the adherence of the Tsallis entropy to the
large deviation theory has generated considerable interest and
debate within the physics community (see, for example \cite{11,12,13} and
the references therein). In particular, work by La Cour and
Schieve \cite{5} showed the canonical probability densities obtained
from Tsallis maximum entropy principle to be generally inconsistent with
the large deviation theory.  The absence of a probabilistic
justification for the Tsallis maximum  entropy principle has
hitherto constituted a significant drawback to the study and
utilization of generalized statistics in such formulations.

This paper attempts to reconcile canonical probability distributions obtained
from the Tsallis maximum entropy principle with the large
deviation theory, based on a procedure that radically differs from that employed in Ref. \cite{5}, and utilizing physically realistic
expectation values for internal energies (hereinafter referred to as energy expectations).  Nonadditive statistics has employed a number of forms in which expectations may be defined. Prominent among these are the linear
constraints originally employed by Tsallis \cite{1} (also known as \textit{normal averages}) of the form: $ \left\langle A
\right\rangle = \sum\limits_i {p_i } A_i $, the Curado-Tsallis (C-T)
constraints \cite{14} of the form:  $ \left\langle A \right\rangle _q  = \sum\limits_i {p_i^q } A_i  \ $, and the normalized
Tsallis-Mendes-Plastino (TMP) constraints \cite{15} (also known as
\textit{$q$-averages}) of the form:  $ \left\langle {\left\langle A
\right\rangle } \right\rangle _q  = \sum\limits_i {\frac{{p_i^q
}}{{\sum\limits_i {p_i^q } }}A_i } \ $. Note that in this paper,
$<\bullet>$ denotes an expectation.  Of
these three methods to describe expectations, the most commonly
employed by Tsallis-practitioners is the TMP-one.

The originally employed normal averages constraints were abandoned
because of  serious difficulties in evaluating the partition
function. The C-T constraints were replaced by the TMP constraints
because they entail the strange relation  $\langle 1 \rangle_q
\neq 1$. Recent works by Abe \cite{16,17,18} suggest that in generalized
statistics expectations defined in terms of normal averages, in
contrast to those defined by $q$-averages, are consistent with the
generalized H-theorem and the generalized \textit{Stosszahlansatz}
(molecular chaos hypothesis).  This resulted in the re-formulation of the variational perturbation approximation in generalized statistics \cite{19}.

This paper employs the physically
tenable normal averages energy expectations, and specifies the
dual Tsallis entropy as the measure of uncertainty. Specifically, it is of essence to utilize the additive duality,
defined by re-parameterizing the nonadditive $q$-parameter by
specifying: $q\rightarrow 2-q=q^*$ \cite{20,21,22}, when employing normal averages energy expectations.  In this instance "$\rightarrow $" denotes a re-parameterization of the nonadditive
parameter, and is not a limit.  Application of the additive
duality to the Tsallis entropy \cite{1}, yields the dual
Tsallis entropy:
\begin{equation}
S_{q^ *  } \left[ {p_i } \right] =  - \sum\limits_i {p_i \ln _{q^ *  } } p_i,
\end{equation}
where the $q^*$-logarithm is defined as: $ \ln _{q^ *  } x =
\frac{{x^{1 - q^ *  }  - 1}}{{1 - q^ *  }}$ \cite{23}. It is readily
seen that in the limit $q^*\rightarrow 1$, the dual Tsallis
entropy (1) tends to the Shannon entropy.

It is noteworthy to mention that the additive duality was recently employed to successfully demonstrate that the dual generalized Kullback-Leibler divergence is a scaled Bregman divergence \cite{24,25}. This paper derives the necessary conditions to reconcile the dual Tsallis maximum  entropy
principle with the asymptotic frequencies obtained from large
deviation theory (i.e. the law of large numbers), employing normal averages energy expectations. These necessary conditions, which enforce the criterion that the canonical probabilities satisfying the $q^*$-maximum entropy principle exactly coincide with those satisfying the Shannon-Jaynes theory, cannot obtained through the analysis in Ref. \cite{5}. The $q^*$-maximum entropy principle is shown \emph{not} to explicitly adhere to the large deviation theory.

However, the necessary conditions are demonstrated to constitute an \emph{invertible mapping} between:  $(i)$ a canonical ensemble satisfying the $q^*$-maximum entropy principle for energy-eigenvalues $\varepsilon_i^*; i=1,...,m\geq 3$ and parameterized by $\beta^* \in [0,\infty]$, and, $(ii)$ a canonical ensemble satisfying the Shannon-Jaynes maximum entropy theory for energy-eigenvalues $\varepsilon_i; i=1,...,m\geq 3$ and parameterized by $\beta$.  The analysis and implications of the invertible mapping, and its role in implicitly reconciling the $q^*$-maximum entropy principle with the large deviation theory, is discussed in Sections V-VII of this paper.  Note that in Sections V-VII of this paper, the terms \textit{necessary conditions} and \textit{invertible mapping} are employed interchangeably. Numerical examples
for exemplary cases are provided.
\section{Dual Tsallis maximum entropy principle}
Following the procedure suggested by Ferri, Martinez, and Plastino
\cite{26},  the canonical probability distribution that maximizes the
dual Tsallis entropy for the energy-eigenvalues $\varepsilon_i=\{\varepsilon_i,...,\varepsilon_m\},m\geq 3$ subject to the constraint:
\begin{equation}
\sum\limits_{i = 1}^m {p_i \varepsilon _i }  = u,
\end{equation}
and, the normalization constraint: $\sum\limits_{i = 1}^m {p_i }  = 1$ is:
\begin{equation}
p_i  = \frac{{\left[ {1 - \left( {1 - q^*  } \right)\beta ^ *  \varepsilon _i } \right]^{\frac{1}{{1 - q^ *  }}} }}{{Z\left( {\beta ^ *  }
 \right)}} = \frac{{\exp _{q^ *  } \left[ { - \beta ^ *  \varepsilon _i } \right]}}{{Z\left( {\beta ^ *  } \right)}},
\end{equation}
where: $\left[ {1 + \left( {1 - q^ *  } \right)x} \right]^{\frac{1}{{1 - q^ *  }}}  = \exp _{q^ *  } x$ \cite{23}.  Here:
\begin{equation}
\begin{array}{l}
 \beta ^ *   = \frac{\beta_{q^*} }{{\left( {2 - q^ *  } \right)\beta_{q^*}u + \left( {1 - q^ *  } \right)\aleph _{q^ *  } }};\aleph _{q^ *  }
   = \sum\limits_i {p_i^{2 - q^ *  } } , \\
 \\

Z\left( {\beta ^ *  } \right) = \sum\limits_i {\exp _{q^ *  } \left[ { - \beta ^ *  \varepsilon _i } \right]}. \\
 \end{array}
\end{equation}
Here, $\beta_{q^*}$ in (4) is the Lagrange multiplier for the
internal energy constraint (2), and $\beta^*$ is referred to as
the \textit{effective inverse temperature}.  As $q^* \rightarrow
1$, $\beta_{q^*}\rightarrow \beta$, where $\beta$ is the
Boltzmann-Gibbs inverse thermodynamic temperature.  Going by the
prescription of Ref. \cite{26}, instead of canonically evaluating the
self-referential expression for $\beta^*$, a
\textit{parametric approach} is adopted by \textit{a-priori}
specifying $\beta^* \in [0,\infty]$.

Consider a sampling distribution $\mu$.  The distribution of
frequencies obtained from  the random samples $x_1,...,x_n$ tends
to $\mu$ as $n\rightarrow \infty$ \cite{5,9}.  Let
$f_{n,i}(x_1,...,x_n)$ be the observed frequency of the discrete
energy-eigenvalues $\varepsilon_i$ in the sample $x_1,...,x_n$.
Thus, (2) may be stated as:
\begin{equation}
\sum\limits_{i = 1}^m {(\varepsilon _i-u) f_{n,i}(x_1,...,x_n) }  = 0.
\end{equation}
It will be demonstrated herein that random samples drawn from $\mu$  and  satisfying
(5) can never give rise to adequate empirical distributions that converge
to the dual Tsallis prediction of (3).  This is highlighted in Sections IV  and VII of this paper.
The negative result has prompted the establishment of an \emph{implicit} adherence of the $q^*$-maximum entropy principle to the large deviation theory, facilitated by an invertible mapping described in Sections V-VII of this paper.

\section{Conditional convergence under constraints}

Herein, the convergence
in probability of the empirical frequencies $f_n =
(f_{n,1}, . . . , f_{n,m})$, where $f_n$ is a random vector with domain
$\left\{ {\varepsilon _1 ,...,\varepsilon _m } \right\}^n$ taking values in the convex set $
\left\{ {p \in \Re ^m ;p_i  > 0;\sum\nolimits_{i = 1}^m {p_i }  = 1} \right\}$, is analyzed.  In an unconstrained setting, Sanov's theorem \cite{5,9}
yields the large deviation
rate function for this convergence to be just the negative
of the Boltzmann-Gibbs-Shannon entropy and a constant:
\begin{equation}
I_\mu \left( p \right)   =  - S_{q^ *   = 1} \left( p \right) - \ln m,
\end{equation}
Imposition of additional constraints on $f_n$, results in the
asymptotic value changing from $\mu$ to a new distribution
which minimizes $I_{\mu}$ under the added restrictions \cite{5,9,10,27}.  For example, imposing the normal averages expectation:
\begin{equation}
\sum\limits_{i = 1}^m {\varepsilon _i f_{n,i}(x_1,...,x_n) }  = u,
\end{equation}
on the sample mean results in an asymptotic distribution which is distinct from $\mu$, and which satisfies $
P_i  \propto e^{ - \beta \varepsilon _i }$
 , where $\beta$ is the Boltzmann-Gibbs inverse thermodynamic temperature. It is ascertained that:
\begin{equation}
\sum\limits_{i = 1}^m {\varepsilon _i P_i  = u.}
\end{equation}

Imposing the condition in (5) yields an
asymptotic canonical distribution which minimizes $I_{\mu}$ and maximizes $S_{q^ *   = 1}$ subject to the normal averages energy expectations
(2):
\begin{equation}
P_i  = \frac{{\exp \left[ { - \beta \varepsilon _i } \right]}}{{Z\left( \beta  \right)}}.
\end{equation}

\section{Difference between canonical distributions}

The leitmotif of this paper is to derive
necessary conditions that allow for agreement between the
canonical probabilities $p_i$ and $P_i$ when $m \geq 3$. To demonstrate this explicitly, the necessary conditions are derived such that for  \textit{a-priori} specified $\beta$ and energy eigenvalues $\varepsilon_i=\{\varepsilon_1,...,\varepsilon_m\}$ and specifying $p_i=P_i$, results in the coincidence of the solutions of (3) and (9) for a state-independent $\beta^*\in[0,\infty]$.
The difference between the distributions given by (3)  and (9) is:
\begin{equation}
d_i \left( \beta  \right) = \frac{{\exp \left[ { - \beta \varepsilon _i } \right]}}{{Z\left( \beta  \right)}} - \frac{{\exp _{q^ *  } \left[ { - \beta
^ *  \varepsilon _i } \right]}}{{Z\left( {\beta ^ *  } \right)}},i=1,...,m,\
\end{equation}
where:
\begin{equation}
\begin{array}{l}
 Z\left( \beta  \right) = \sum\limits_{i = 1}^m {\exp \left[ { - \beta \varepsilon _i } \right],}  \\
 \\
 and, \\
 \\
 Z\left( {\beta ^ *  } \right) = \sum\limits_{i = 1}^m {\exp _{q^ *  } \left[ { - \beta ^ *  \varepsilon _i } \right]}.  \\
 \end{array}
\end{equation}
The necessary conditions are obtained by enforcing the condition:
\begin{equation}
d_i(\beta)=0;i=1,...,m.
\end{equation}
Here, (12) tacitly mandates that the canonical distributions (3) and (9) exactly coincide.

Enforcing the condition in (12), yields:
\begin{equation}
 Z^ *  \exp \left[ { - \beta \varepsilon _i } \right] = \exp _{q^ *  } \left[ { - \beta ^ *  \varepsilon _i } \right];i=1,...,m, \\
\end{equation}
where: $Z^*=\frac{{Z\left( {\beta ^ *  } \right)}}{{Z\left( \beta  \right)}}>0$.  From (13), it is immediately evident that values of
$\varepsilon_i=0$ result in $Z(\beta^*)=Z(\beta)$, which is unphysical.

As will be demonstrated in Section VII of this paper, the mapping of (3) onto (9) \emph{cannot} be achieved for any \textit{a-priori} state-independent values of $\beta^*$ and energy-eigenvalues $\varepsilon_i, i=1,...,m,m\geq 3$.  This tacitly implies that canonical probabilities satisfying the $q^*$-maximum entropy principle can \emph{never} be made to \emph{explicitly} adhere to the large deviation theory.
\section{the invertible mapping}
To ameliorate this intractable situation, an \emph{implicit} procedure is adopted to allow for the adherence of canonical probabilities satisfying the $q^*$-maximum entropy principle to the large deviation theory.  This \emph{implicit} adherence is accomplished by the introduction of an \emph{invertible mapping}.

Consider a sampling distribution $\mu^*$.  The distribution of
frequencies obtained from  the random samples $x_1^*,...,x_n^*$.  Let
$f_{n,i}^*(x_1^*,...,x_n^*)$ be the observed frequency of the discrete
energy-eigenvalues $\varepsilon_i^*$ in the sample $x_1^*,...,x_n^*$.  Here, $f_n^* =
(f_{n,1}^*, . . . , f_{n,m}^*)$, where $f_n^*$ is a random vector with domain
$\left\{ {\varepsilon _1^* ,...,\varepsilon _m^* } \right\}^n$ taking values in the convex set $
\left\{ {p^* \in \Re ^m ;p_i^*  > 0;\sum\nolimits_{i = 1}^m {p_i^* }  = 1} \right\}$.

The objective of the invertible mapping is to relate:  $(i.)$ a canonical ensemble satisfying the $q^*$-maximum entropy principle for energy-eigenvalues $\varepsilon_i^*$ ,with observed frequencies $f_{n,i}^*(x_1^*,...,x_n^*)$, and parameterized by a constant $\beta^* \in [0,\infty]$, and, $(ii.)$ a canonical ensemble satisfying the Shannon-Jaynes maximum entropy theory for energy-eigenvalues $\varepsilon_i$, with observed frequencies $f_{n,i}(x_1,...,x_n)$, and parameterized by a constant $\beta$.

For the energy-eigenvalues $\varepsilon_i^*=\{\varepsilon_i^*,...,\varepsilon_m^*\},m\geq 3$, (2) and (5) are re-specified as:
\begin{equation}
\sum\limits_{i = 1}^m {(\varepsilon _i^*-u) f_{n,i}^*(x_1^*,...,x_n^*) }  = \sum\limits_{i = 1}^m {(\varepsilon _i^*-u)p_i^*  }  = 0.
\end{equation}

Here, (3)  is re-defined as:
\begin{equation}
\begin{array}{l}
p_i^*  = \frac{{\left[ {1 - \left( {1 - q^*  } \right)\beta ^ *  \varepsilon _i^* } \right]^{\frac{1}{{1 - q^ *  }}} }}{{Z_{\mu^*}\left( {\beta ^ *  }
 \right)}} = \frac{{\exp _{q^ *  } \left[ { - \beta ^ *  \varepsilon _i^* } \right]}}{{Z_{\mu^*}\left( {\beta ^ *  } \right)}},\\
\\
where,\\
\\
Z_{\mu^*}\left( {\beta ^ *  } \right) = \sum\limits_i {\exp _{q^ *  } \left[ { - \beta ^ *  \varepsilon _i^* } \right]}. \\
\end{array}
\end{equation}

The difference between the distributions given by (15) and (9), is re-defined as:
\begin{equation}
\begin{array}{l}
d_i \left( \beta  \right) = \frac{{\exp \left[ { - \beta \varepsilon _i } \right]}}{{Z\left( \beta  \right)}} - \frac{{\exp _{q^ *  } \left[ { - \beta
^ *  \varepsilon _i^* } \right]}}{{Z_{\mu^*}\left( {\beta ^ *  } \right)}},i=1,...,m,\\
\\
where:
\\

 Z\left( \beta  \right) = \sum\limits_{i = 1}^m {\exp \left[ { - \beta \varepsilon _i } \right],}  \\
 \\
 and, \\
 \\
 Z_{\mu^*}\left( {\beta ^ *  } \right) = \sum\limits_{i = 1}^m {\exp _{q^ *  } \left[ { - \beta ^ *  \varepsilon _i^* } \right]}.  \\
 \end{array}
\end{equation}

The necessary conditions (13) now acquire the form:
\begin{equation}
 Z_{\mu^*}^ *  \exp \left[ { - \beta \varepsilon _i } \right] = \exp _{q^ *  } \left[ { - \beta ^ *  \varepsilon _i^* } \right];i=1,...,m, \\
\end{equation}
where: $Z_{\mu^*}^*=\frac{{Z_{\mu^*}\left( {\beta ^ *  } \right)}}{{Z\left( \beta  \right)}}>0$.  Note that all quantities with the subscript ${\mu^*}$, are evaluated for the energy-eigenvalues $\varepsilon_i^*=\{\varepsilon_i^*,...,\varepsilon_m^* \}$, and are parameterized by $\beta^*$.

By definition:
\begin{equation}
\exp \left[ { - \beta \varepsilon _i } \right] = Z\left( \beta  \right) - \sum\limits_{k \ne i} {\exp \left[ { - \beta \varepsilon _k } \right]}  =
Z\left( \beta  \right)-\Phi _i. \\
\end{equation}

Substituting (18) into (17), and taking the $q^*$-logarithm on both sides, yields the \emph{invertible mapping}:
\begin{equation}
\tau_i^*=\beta^* \varepsilon_i^*   =  -  \ln_{q^*}\left[ {Z_{\mu^*}^*  \left( {Z\left( \beta  \right) - \Phi _i } \right)} \right];i=1,...,m.
\end{equation}

The invertible mapping (19) transforms: $(i.)$ canonical probabilities satisfying the $q^*$-maximum entropy principle for energy-eigenvalues $\varepsilon_i^*$ , and parameterized by a state-independent $\beta^* \in [0,\infty]$, into, $(ii.)$ canonical probabilities satisfying the Shannon-Jaynes maximum entropy theory for energy-eigenvalues $\varepsilon_i$, and parameterized by a constant $\beta$.
Specifically, Eq. (19) invertibly transforms:
\begin{equation}
\frac{{\exp _{q^ *  } \left[ { - \beta
^ *  \varepsilon _i^* } \right]}}{{Z_{\mu^*}\left( {\beta ^ *  } \right)}} \
 \leftrightarrow \frac{{\exp \left[ { - \beta \varepsilon _i } \right]}}{{Z\left( \beta  \right)}}; i=1,...,m.
\end{equation}

The objective of the invertible mapping (19) is to evaluate  $q^*$-canonical probabilities $p_i^*$ from (15) using the energy-eigenvalues $\varepsilon_i^*$ parameterized by $\beta^*$, and, transform them into Boltzmann-Gibbs canonical probabilities $P_i$ defined in (9) in terms of energy-eigenvalues $\varepsilon_i$ parameterized by $\beta$.  Here, (19) also facilitates the inverse transformation.  The leitmotif for this transform is to overcome the formidable obstacles encountered when attempting to explicitly show adherence of the $q^*$-maximum entropy principle to the large deviation theory.  

Thus, first canonical probabilities $p_i^*$ are obtained and then transformed into the Boltzmann-Gibbs form $P_i$ defined by (9), thereby trivially satisfying (16).  In this
analysis, (18) is introduced so as to express $\tau_i^*$ in terms of
both the canonical partition function ($Z(\beta)$) and a subset of
of it ($\Phi_i$) that accounts for contributions of discrete
eigenvalues $\varepsilon_k$, $k\neq i$.

\section{Utility of the invertible mapping}

Eq. (9) yields:
\begin{equation}
\begin{array}{l}
 P_i  = \frac{{\exp \left[ { - \beta \varepsilon_i } \right]}}{{Z\left( \beta  \right)}}, \\
 \\
 {\rm with} \\
 \\
 Z\left( \beta  \right) = \sum\limits_{j = 1}^m {\exp \left[ { - \beta \varepsilon _j } \right]} , \\
 \end{array}
\end{equation}
where $i,j=1,...,m$.  Taking logarithms in the first relation of (21) yields:
\begin{equation}
\ln P_i  =  - \beta \varepsilon _i  - \ln Z\left( \beta  \right).
\end{equation}

The values of $\beta$ and $\ln Z(\beta)$ remain constant $\forall$
$\varepsilon_i$ and $\varepsilon_j$, and consequently, $\forall$
$P_i$ and $P_j$;$j\neq i$, $i,j=1,...,m$.  Thus, (22) may also be
expressed as:
\begin{equation}
\begin{array}{l}
 \ln P_j  =  - \beta \varepsilon _j  - \ln Z\left( \beta  \right) \\
  \\
  \Rightarrow \beta  =  - \frac{{\left[ {\ln P_j  + \ln Z\left( \beta  \right)} \right]}}{{\varepsilon _j }}. \\
 \end{array}
\end{equation}
Substituting (25) back into the first relation of (23) and
re-arranging leads to:
\begin{equation}
 \ln Z\left( \beta  \right) = \frac{{\varepsilon _i \ln P_j  - \varepsilon _j \ln P_i }}{{\varepsilon _j  - \varepsilon _i }};i\neq j, i,j=1,...,m.
 \\
\end{equation}

Here, Eq. (24) is readily  satisfied for $q^*=1$ (Boltzmann-Gibbs canonical probabilities); $\forall i\neq j, i,j=1,...,m,m\geq 3$.  Hence the utility of the invertible mapping defined by Eq. (19), which transforms a canonical probability distribution $p_i^*$ satisfying the $q^*$-maximum entropy principle for energy-eigenvalues $\varepsilon_i^*, i=1,...,m$ and parameterized by a constant (state-independent) $\beta^*$, to an exactly equivalent Boltzmann-Gibbs canonical probability distribution $p_i$ satisfying the Shannon-Jaynes maximum entropy theory for energy-eigenvalues $\varepsilon_i, i=1,...,m$ and parameterized by a constant $\beta$, and \emph{vice-versa}.

The analysis in this Section treats two separate and distinct state indices $i$ and $j$; $i\neq j, i,j=1,...,m$.  Thus, (19) is re-stated as:
\begin{equation}
\tau_j^*=\beta ^ *\varepsilon_j^*   =  -  \ln_{q^*}\left[ {Z_{\mu^*}^ *  \left( {Z\left( \beta  \right) - \Phi _j } \right)} \right];j=1,...,m,
\end{equation}
where, $\beta^* \in [0,\infty], \forall j$.

The LHS of (24) is required to be the same $\forall i,j, i\neq j$. Invoking (19), (15) acquires the form:
\begin{equation}
\begin{array}{l}
 p_i^*  = \frac{{\exp _{q^ *  } \left[ { - \tau_i ^ *} \right]}}{{Z_{\mu^*}\left( {\beta ^ *  } \right)}} = \frac{{Z_{\mu^*}^ *  \left( {Z\left( \beta
 \right) - \Phi _i } \right)}}{{Z_{\mu^*}\left( {\beta ^ *  } \right)}}, \\
 \\
 and, \\
 \\
 p_j^* = \frac{{\exp _{q^ *  } \left[ { - \tau_j ^ * } \right]}}{{Z_{\mu^*}\left( {\beta ^ *  } \right)}} = \frac{{Z_{\mu}^ *  \left( {Z\left( \beta
 \right) - \Phi _j } \right)}}{{Z_{\mu^*}\left( {\beta ^ *  } \right)}}, \\
 \end{array}
\end{equation}
$j\neq i;i,j=1,...,m$.    Utilizing the relation: $Z_{\mu^*}^ *   = \frac{{Z_{\mu^*}\left( {\beta ^ *  } \right)}}{{Z\left( \beta  \right)}}$, (26) yields the Boltzmann-Gibbs canonical probability distributions:
\begin{equation}
\begin{array}{l}
 p_i  = \frac{{\left( {Z\left( \beta
 \right) - \Phi _i } \right)}}{{Z\left( {\beta   } \right)}}, \\
 \\
 and, \\
 \\
 p_j  = \frac{{\left( {Z\left( \beta
 \right) - \Phi _j } \right)}}{{Z\left( {\beta  } \right)}}, \\
 \end{array}
\end{equation}
$j\neq i;i,j=1,...,m$. Substituting (27) into (24), yields:
\begin{equation}
\begin{array}{l}
 \ln Z\left( \beta  \right)\mathop   \frac{{\varepsilon _i \ln \left[ {\frac{{Z\left( \beta  \right) - \Phi _j }}{{Z\left(
 \beta  \right)}}} \right] - \varepsilon _j \ln \left[ {\frac{{Z\left( \beta  \right) - \Phi _i }}{{Z\left( \beta  \right)}}} \right]}}{{\delta
 \varepsilon _{ji} }}\\
 \\
= \frac{{\varepsilon _i \ln \left[ {Z\left( \beta  \right) - \Phi _j } \right] - \varepsilon _i \ln Z\left( \beta  \right) - \varepsilon _j \ln
  \left[ {Z\left( \beta  \right) - \Phi _i } \right] + \varepsilon _j \ln Z\left( \beta  \right)}}{{\delta \varepsilon _{ji} }} \\
  \\
  = \frac{{\varepsilon _i \ln \left[ {Z\left( \beta  \right) - \Phi _j } \right] - \varepsilon _j \ln \left[ {Z\left( \beta  \right) - \Phi _i }
  \right]}}{{\delta \varepsilon _{ji} }} + \ln Z\left( \beta  \right); \\
\\
\Rightarrow \ln Z(\beta)= \frac{{\varepsilon _j \ln \Phi _i  - \varepsilon _i \ln \Phi _j }}{{\delta \varepsilon _{ji}
  }};
 i,j=1,...,m,i\neq j, \\
 \end{array}
\end{equation}
where:  $\delta\varepsilon_{ji}=\varepsilon_j-\varepsilon_i$. Expanding (28) with the aid of (18) yields:
\begin{equation}
\begin{array}{l}
 \delta \varepsilon _{ji} \ln Z\left( \beta  \right) \\
 \\
  = \delta \varepsilon _{ji} \ln Z\left( \beta  \right) + \varepsilon _j \beta \varepsilon _i  - \varepsilon _i \beta \varepsilon _j=\delta \varepsilon _{ji}\ln Z\left( \beta  \right)\\
  \\
\Rightarrow \ln Z\left( \beta  \right)=\ln Z\left( \beta  \right);i,j=1,...,m,i\neq j. \\
 \end{array}
\end{equation}

Here, (29)
tacitly demonstrates the consistency of the LHS of (24) $\forall i,j=1,...,m;i \neq j$.  It is important to note that the identical results to those described in (29) may be obtained by applying the theory described in this paper to the model described in Ref. \cite{5}.

\section{Implementation of the invertible mapping}

The necessary conditions (19) are obtained by enforcing the requirement that the distance between the canonical probability distributions (9) and (15), given by (16), vanishes.  This is mandated by Eq. (12) of this paper. The necessary conditions (19)  may thus be interpreted as an invertible mapping that transforms (15) into (9).  In order that the solutions of (9) coincide with those of (15) for a constant $\beta$, the effective inverse temperature $\beta^*$ has to be state-independent $\forall \varepsilon_i^*$.

This requirement stems from one of the fundamental tenets of statistical physics that in a given canonical ensemble, the energy-eigenvalues constitute a spectrum,  but the
thermodynamic temperature (or within the context of this analysis, the effective inverse temperature $\beta^*$) is fixed, is .
Taking the logarithm of (9) for \textit{a-priori} specified $\beta$ and energy-eigenvalues $\varepsilon_i, i=1,...,m$, yields:
\begin{equation}
\ln P_i  =  - \beta \varepsilon _i  - \ln Z\left( \beta  \right); i=1,...,m.
\end{equation}
For $\beta=0.5$, $\varepsilon_i=\{1.0,2.0, 3.0\}; i=1, 2, 3$ versus $\ln P_i$, obtained from (30), is a straight line, as is demonstrated in Fig. 1.  The value of corresponding canonical partition function, $Z(\beta)$=1.1975.
Likewise, taking the logarithm of (15), yields:
\begin{equation}
\ln p_i^*  = \ln \left\{ {\exp _{q^ *  } \left[ { - \beta^ *  \varepsilon _i^* } \right]} \right\} - \ln Z_{\mu^*}\left( {\beta^ *  } \right); i=1,...,m.
\end{equation}
For a constant $\beta^* \in [0,\infty]$, and energy-eigenvalues $\varepsilon_i^*=\varepsilon_i$, the solutions of (31) \emph{cannot} coincide with those of (30) for the same canonical ensemble with constant $\beta^*$, except perhaps for $m\leq 2$.

For example, even for a simple set of three energy-eigenvalues:$\{\varepsilon_1,\varepsilon_2,\varepsilon_3\}$, such a coincidence of canonical probabilities for constant $\beta$ and $\beta^*$, requires that the following overly prohibitive conditions be simultaneously satisfied:
\begin{equation}
\begin{array}{l}
 \beta ^ *   = -\frac{{\ln _{q^ *  } \left\{ {Z^ *  \exp \left[ { - \beta \varepsilon _1 } \right]} \right\}}}{\varepsilon _1 },\\
\\
 = -\frac{{\ln _{q^ *  } \left\{ {Z^ *  \exp \left[ { - \beta \varepsilon _2 } \right]} \right\}}}{\varepsilon _2 }, \\

 \\
 = -\frac{{\ln _{q^ *  } \left\{ {Z^ *  \exp \left[ { - \beta \varepsilon _3 } \right]} \right\}}}{\varepsilon _3 }.\\
 \end{array}
\end{equation}
The derivation of (32) is detailed in the Appendix of this paper.

It is, however, readily demonstrated that the solutions of (31) and (30) will coincide, using the invertible mapping (19).  Setting $Z_{\mu^*}^*=0.9$ in (19) and specifying $\beta^*=1.5$, yields the energy-eigenvalues: $\varepsilon_i^*=\{0.4713, 0.9839,  1.6420 \}$.  Here, $Z_{\mu^*}(\beta^*)=1.0778$ for $q^*=1.5$.  Fig. 2 depicts $\varepsilon_i^*; i=1,2,3$ versus $\ln p_i^*$.  Note that the above mentioned numerical simulation results obey Eq. (32), which is gainfully utilized, by re-stating it in the form:
\begin{equation}
\begin{array}{l}
 \beta ^ *   = -\frac{{\ln _{q^ *  } \left\{ {Z_{\mu^*}^ *  \exp \left[ { - \beta \varepsilon _1 } \right]} \right\}}}{\varepsilon _1^* },\\
\\
 = -\frac{{\ln _{q^ *  } \left\{ {Z_{\mu^*}^ *  \exp \left[ { - \beta \varepsilon _2 } \right]} \right\}}}{\varepsilon _2^* }, \\

 \\
 = -\frac{{\ln _{q^ *  } \left\{ {Z_{\mu^*}^ *  \exp \left[ { - \beta \varepsilon _3 } \right]} \right\}}}{\varepsilon _3^* },\\
\\
...\\
\\
= -\frac{{\ln _{q^ *  } \left\{ {Z_{\mu^*}^ *  \exp \left[ { - \beta \varepsilon _m } \right]} \right\}}}{\varepsilon _m^* }.
 \end{array}
\end{equation}

Analytically, the invertible mapping (19) may be readily shown to facilitate the coincidence of (31) with (30) $\forall i=1,...,m; m\geq 3$.  Substituting (19) into (31), results in:
\begin{equation}
\begin{array}{l}
 \ln p_i^*  = \ln \exp _{q^ *  } \left\{ {\ln _{q^ *  } \left[ {Z_{\mu^*}^ *  \left( {Z\left( \beta  \right) - \Phi _i } \right)} \right]} \right\} - \ln Z_{\mu^*}\left( {\beta^ *  } \right)\\
 \\
  = \ln \left[ {Z_{\mu^*}^ *  \left( {Z\left( \beta  \right) - \Phi _i } \right)} \right] - \ln Z_{\mu^*}\left( {\beta^ *  } \right) \\
  \\
  = \ln Z_{\mu^*}^ *   + \ln \left( {Z\left( \beta  \right) - \Phi _i } \right) - \ln Z_{\mu^*}\left( {\beta^ *  } \right) \\
\\
\mathop  = \limits^{\left( a \right)}  - \beta \varepsilon _i  - \ln Z\left( \beta  \right) \Rightarrow {\rm Eq. (30) ~recovered!}
\end{array}
\end{equation}
Here, $(a)$ denotes substituting the expressions: $Z_{\mu^*}^ *   = \frac{{Z_{\mu^*}\left( {\beta^ *  } \right)}}{{Z\left( \beta  \right)}}$ and (18), and re-arranging.  Thus, (30) is seamlessly recovered from (31) with the aid of the invertible mapping (19).  The calculation in (34) does not specify any restriction on the value of $m$.

In the above discussion and the numerical simulations depicted in Fig. 2, $(i)$ $\beta$ and energy-eigenvalues $\varepsilon_i$ are given \emph{a-priori}, $(ii)$ $Z_{\mu^*}^*$, $\beta^*$, and the nonadditive parameter $q^*$ are arbitrarily specified, and $(iii)$ the energy-eigenvalues $\varepsilon_i^*$ are derived from Eq. (19), while ensuring that Eq. (33) is satisfied.   It is important to note that the analysis presented herein contains a number of parameters.  It is thus imperative to ensure that the fidelity of the invertible mapping (Eq. (19)), which nonlinearly relates $Z_{\mu^*}^*$, $\beta^*$, $\beta$, $\varepsilon_i^*$, and $\varepsilon_i$ is retained, while simultaneously satisfying Eq. (33).  A comprehensive multi-parameter study of the invertible mapping is beyond the scope of this paper, and will be presented elsewhere.  

\section{Discussions and Conclusions}

This paper demonstrates that a given canonical ensemble satisfying the $q^*$-maximum entropy principle does not adhere to the large deviation theory.  A unique and robust invertible mapping is tacitly demonstrated to facilitate the \emph{implicit} adherence of a given canonical ensemble satisfying the $q^*$-maximum entropy principle, to the large deviation theory.

This is accomplished by: $(i)$  Obtaining canonical probabilities $p_i^*$ defined by Eq. (15), satisfying the $q^*$-maximum entropy principle for energy-eigenvalues $\varepsilon_i^*=\{\varepsilon_1^*,...,\varepsilon_m^* \}; m\geq 3$, and parameterized by  a  state-independent effective inverse temperature $\beta^* \in [0,\infty]$, $(ii)$ Utilizing the invertible mapping described by Eq. (19) to obtain canonical probabilities satisfying the Shannon-Jaynes maximum entropy theory for energy-eigenvalues $\varepsilon_i=\{\varepsilon_1,...,\varepsilon_m\}; m\geq 3$, and parameterized by a constant inverse thermodynamic temperature $\beta$, and, $(iii)$ demonstrating that the difference between the canonical probabilities (Eq. (16)) is trivially satisfied, thereby proving an \emph{implicit} adherence of the $q^*$-maximum entropy principle to the large deviation theory.

This is specifically accomplished by mapping a canonical ensemble that satisfies $q^*$-maximum entropy principle but does not adhere to the large deviation theory, onto, a canonical ensemble that satisfies the Shannon-Jaynes theory whose foundations are intimately intertwined to the large deviation theory \cite{9,10,27}.  Note that the invertible mapping (Eq. (19)) seamlessly allows for the interchange of steps $(i)$ and $(ii)$ described above, depending upon the energy-eigenvalues and other parameters made available.

Numerical results  for exemplary cases have been provided.  The
results presented in this paper constitute a substantial qualitative improvement \textit{vis-\'{a}-vis} those
demonstrated in previous studies, that have attempted to reconcile the
$q$-maximum entropy principle with the large deviation theory. A
comprehensive  analysis of the invertible mapping (Eq.(19)), from
information geometric considerations, is currently being pursued.  This analysis also assesses the sensitivity of the invertible mapping to variations in $Z_{\mu^*}^*$, $\beta^*$, $\beta$, $\varepsilon_i^*$, and $\varepsilon_i$.
The pertinent results  will  be presented elsewhere.

\acknowledgments
RCV was supported by \textit{NSFC} contract
\textit{111017-01-2013}.

\nocite{*}

\bibliography{apssamp}

\begin{thebibliography}{3}
\expandafter\ifx\csname natexlab\endcsname\relax\def\natexlab#1{#1}\fi
\expandafter\ifx\csname bibnamefont\endcsname\relax
  \def\bibnamefont#1{#1}\fi
\expandafter\ifx\csname bibfnamefont\endcsname\relax
  \def\bibfnamefont#1{#1}\fi
\expandafter\ifx\csname citenamefont\endcsname\relax
  \def\citenamefont#1{#1}\fi
\expandafter\ifx\csname url\endcsname\relax
  \def\url#1{\texttt{#1}}\fi
\expandafter\ifx\csname urlprefix\endcsname\relax\def\urlprefix{URL }\fi
\providecommand{\bibinfo}[2]{#2}
\providecommand{\eprint}[2][]{\url{#2}}

\bibitem[1]{1}
\bibinfo{author}{\bibfnamefont{C.} \bibnamefont{Tsallis}},
  \bibinfo{book}{\textit{Introduction to Nonextensive Statistical Mechanics: Approaching a Complex World}}, (Springer-Verlag, New York, 2009).


\bibitem[2]{2}
\bibinfo{author}{\bibfnamefont{D. H.}~\bibnamefont{Zanette}}
\bibnamefont{and}
  \bibinfo{author}{\bibfnamefont{M. M.}~\bibnamefont{Montemurro}},
  \bibinfo{journal}{Phys. Lett.
A} \textbf{\bibinfo{volume}{316}},
  \bibinfo{pages}{194} (\bibinfo{year}{2003}).

\bibitem[3]{3}
\bibinfo{author}{\bibfnamefont{D. H.}~\bibnamefont{Zanette}}
\bibnamefont{and}
  \bibinfo{author}{\bibfnamefont{M. M.}~\bibnamefont{Montemurro}},
  \bibinfo{journal}{Phys. Lett.
A} \textbf{\bibinfo{volume}{324}},
  \bibinfo{pages}{383} (\bibinfo{year}{2004}).

\bibitem[4]{4}
\bibinfo{author}{\bibfnamefont{E.}~\bibnamefont{Vives}}
\bibnamefont{and}
  \bibinfo{author}{\bibfnamefont{A.}~\bibnamefont{Planes}},
  \bibinfo{journal}{Phys. Rev. Lett.} \textbf{\bibinfo{volume}{88}},  \bibinfo{pages}{02061} (\bibinfo{year}{2002}).

\bibitem[5]{5}
\bibinfo{author}{\bibfnamefont{B.~R.}~\bibnamefont{La Cour}}
\bibnamefont{and}
  \bibinfo{author}{\bibfnamefont{W.~C.}~\bibnamefont{Schieve}},
  \bibinfo{journal}{Phys. Rev. E} \textbf{\bibinfo{volume}{62}},
  \bibinfo{pages}{7494} (\bibinfo{year}{2000}).

\bibitem[6]{6}
\bibinfo{author}{\bibfnamefont{F.}~\bibnamefont{Bouchet}},
  \bibinfo{author}{\bibfnamefont{T.}~\bibnamefont{Dauxois}}, \bibnamefont{and}
  \bibinfo{author}{\bibfnamefont{S.}~\bibnamefont{Ruffo}},
  \bibinfo{journal}{Europhys. News} \textbf{\bibinfo{volume}{37}},
  \bibinfo{pages}{9} (\bibinfo{year}{2008}).

\bibitem[7]{7}
\bibinfo{author}{\bibfnamefont{B.~H.}~\bibnamefont{Lavenda}}
\bibnamefont{and}
  \bibinfo{author}{\bibfnamefont{J.}~\bibnamefont{Dunning-Davies}},
  \bibinfo{journal}{J. Appl. Sci.} \textbf{\bibinfo{volume}{5}},
  \bibinfo{pages}{315} (\bibinfo{year}{2005}).



\bibitem[8]{8}
\bibinfo{author}{\bibfnamefont{M.}~\bibnamefont{Nauenberg}},
  \bibinfo{journal}{Phys. Rev. E} \textbf{\bibinfo{volume}{67}},
  \bibinfo{pages}{036114} (\bibinfo{year}{2003}).

\bibitem[9]{9}
\bibinfo{author}{\bibfnamefont{R.~S.} \bibnamefont{Ellis}},
  \bibinfo{book}{\textit{Entropy, Large Deviations, and Statistical Mechanics}}, (Springer-Verlag, New York, 1985).

\bibitem[10]{10}
\bibinfo{author}{\bibfnamefont{H.}~\bibnamefont{Touchette}},
  \bibinfo{journal}{Phys. Rep.} \textbf{\bibinfo{volume}{478}},
  \bibinfo{pages}{1} (\bibinfo{year}{2009}).


\bibitem[11]{11}
\bibinfo{author}{\bibfnamefont{G.}~\bibnamefont{Ruiz}},
  \bibnamefont{and}
  \bibinfo{author}{\bibfnamefont{C.}~\bibnamefont{Tsallis}},
  \bibinfo{journal}{Phys. Lett. A} \textbf{\bibinfo{volume}{376}},
  \bibinfo{pages}{2451} (\bibinfo{year}{2012}).

\bibitem[12]{12}
\bibinfo{author}{\bibfnamefont{H.}~\bibnamefont{Touchette}},
  \bibinfo{journal}{Phys. Lett. A} \textbf{\bibinfo{volume}{377}},
  \bibinfo{pages}{436} (\bibinfo{year}{2013}).

\bibitem[13]{13}
\bibinfo{author}{\bibfnamefont{G.}~\bibnamefont{Ruiz}},
  \bibnamefont{and}
  \bibinfo{author}{\bibfnamefont{C.}~\bibnamefont{Tsallis}},
  \bibinfo{journal}{Phys. Lett. A} \textbf{\bibinfo{volume}{377}},
  \bibinfo{pages}{377} (\bibinfo{year}{2013}).

\bibitem[14]{14}
\bibinfo{author}{\bibfnamefont{E.~M.~F.}~\bibnamefont{Curado}},
  \bibnamefont{and}
  \bibinfo{author}{\bibfnamefont{C.}~\bibnamefont{Tsallis}},
  \bibinfo{journal}{J. Phys. A: Math Gen.} \textbf{\bibinfo{volume}{24}},
  \bibinfo{pages}{L91} (\bibinfo{year}{1991}).

\bibitem[15]{15}
\bibinfo{author}{\bibfnamefont{C.}~\bibnamefont{Tsallis}},
  \bibinfo{author}{\bibfnamefont{R. S.}~\bibnamefont{Mendes}}, \bibnamefont{and}
  \bibinfo{author}{\bibfnamefont{A.~R.}~\bibnamefont{Plastino}},
  \bibinfo{journal}{Physica A} \textbf{\bibinfo{volume}{261}},
  \bibinfo{pages}{524} (\bibinfo{year}{1998}).

\bibitem[16]{16}
\bibinfo{author}{\bibfnamefont{S.}~\bibnamefont{Abe}},
  \bibinfo{journal}{Phys. Rev. E} \textbf{\bibinfo{volume}{79}},
  \bibinfo{pages}{04116} (\bibinfo{year}{2009}).

\bibitem[17]{17}
\bibinfo{author}{\bibfnamefont{S.}~\bibnamefont{Abe}},
  \bibinfo{journal}{Europhys. Lett.} \textbf{\bibinfo{volume}{84}},
  \bibinfo{pages}{60006} (\bibinfo{year}{2008}).

\bibitem[18]{18}
\bibinfo{author}{\bibfnamefont{S.}~\bibnamefont{Abe}},
  \bibinfo{journal}{J. Stat. Mech.: Th. Expt.} \textbf{\bibinfo{volume}{}},
  \bibinfo{pages}{P07027} (\bibinfo{year}{2009}).
\bibitem[19]{19}
\bibinfo{author}{\bibfnamefont{R. C.}~\bibnamefont{Venkatesan}},
  \bibnamefont{and}
  \bibinfo{author}{\bibfnamefont{A.}~\bibnamefont{Plastino}},
  \bibinfo{journal}{Physica A} \textbf{\bibinfo{volume}{389}},
  \bibinfo{pages}{1159} (\bibinfo{year}{2010});\bibinfo{journal}{Physica A} \textbf{\bibinfo{volume}{389}},
  \bibinfo{pages}{2155} (\bibinfo{year}{2010}).

\bibitem[20]{20}
\bibinfo{author}{\bibfnamefont{T.}~\bibnamefont{Wada}},
  \bibnamefont{and}
  \bibinfo{author}{\bibfnamefont{A. M.}~\bibnamefont{Scarfone}},
  \bibinfo{journal}{Eur. Phys. Jour. B} \textbf{\bibinfo{volume}{47}},
  \bibinfo{pages}{557} (\bibinfo{year}{2005}).

\bibitem[21]{21}
\bibinfo{author}{\bibfnamefont{T.}~\bibnamefont{Wada}},
  \bibnamefont{and}
  \bibinfo{author}{\bibfnamefont{A. M.}~\bibnamefont{Scarfone}},
  \bibinfo{journal}{Phys. Lett. A.} \textbf{\bibinfo{volume}{335}},
  \bibinfo{pages}{351} (\bibinfo{year}{2005}).

\bibitem[22]{22}
\bibinfo{author}{\bibfnamefont{J.}~\bibnamefont{Naudts}},
  \bibinfo{journal}{Physica A} \textbf{\bibinfo{volume}{340}},
  \bibinfo{pages}{32} (\bibinfo{year}{2004}).

\bibitem[23]{23}
\bibinfo{author}{\bibfnamefont{E.}~\bibnamefont{Borges}},
  \bibinfo{journal}{Physica A} \textbf{\bibinfo{volume}{340}},
  \bibinfo{pages}{95} (\bibinfo{year}{2004}).

\bibitem[24]{24}
\bibinfo{author}{\bibfnamefont{R. C.}~\bibnamefont{Venkatesan}},
  \bibnamefont{and}
  \bibinfo{author}{\bibfnamefont{A.}~\bibnamefont{Plastino}},
  \bibinfo{journal}{Physica A} \textbf{\bibinfo{volume}{390}},
  \bibinfo{pages}{2749} (\bibinfo{year}{2011}).

\bibitem[25]{25}
\bibinfo{author}{\bibfnamefont{R. C.}~\bibnamefont{Venkatesan}},
  \bibnamefont{and}
  \bibinfo{author}{\bibfnamefont{A.}~\bibnamefont{Plastino}},
  \bibinfo{journal}{Phys. Lett. A} \textbf{\bibinfo{volume}{375}},
  \bibinfo{pages}{4237} (\bibinfo{year}{2011});\bibinfo{journal}{Phys. Lett. A} \textbf{\bibinfo{volume}{376}},
  \bibinfo{pages}{3470} (\bibinfo{year}{2012}).


\bibitem[26]{26}
\bibinfo{author}{\bibfnamefont{G. L.}~\bibnamefont{Ferri}},
  \bibinfo{author}{\bibfnamefont{S.}~\bibnamefont{Martinez}}, \bibnamefont{and}
  \bibinfo{author}{\bibfnamefont{A.}~\bibnamefont{Plastino}},
  \bibinfo{journal}{J. Stat. Mech.: Th. Expt.} \textbf{\bibinfo{volume}{}},
  \bibinfo{pages}{P04009} (\bibinfo{year}{2005}).

\bibitem[27]{27}
\bibinfo{author}{\bibfnamefont{R. S.}~\bibnamefont{Ellis}},
  \bibinfo{journal}{Physica D} \textbf{\bibinfo{volume}{133}},
  \bibinfo{pages}{106} (\bibinfo{year}{1999}).
\end{thebibliography}

\appendix*
\section{Derivation of Eq. (32)}

For simplicity, a set of three energy-eigenvalues:$\{\varepsilon_1,\varepsilon_2,\varepsilon_3\}$ is chosen, which satisfy the Shannon-Jaynes maximum entropy theory.  From $q$-algebra (Ref. \cite{23}), the following relation is obtained:
\begin{equation}
\begin{array}{l}
 \exp _{q^ *  } \left[ x \right] \otimes _{q^ *  } \exp _{q^ *  } \left[ y \right] = \exp _{q^ *  } \left[ {x + y} \right], \\
\\
where: \\
\\
x \otimes _{q^ *  } y = \left[ {x^{1 - q^ *  }  + y^{1 - q^ *  }  - 1} \right]^{\frac{1}{{1 - q^ *  }}} ;x,y > 0. \\
 \end{array}
\end{equation}
The discrete set of energy-eigenvalues may be denoted as: $\{\varepsilon_1, \varepsilon_1+\delta_1,\varepsilon_1+\delta_2\}$, where $\delta_1,\delta_2 >0 $. Setting:
\begin{equation}
\tau _1^ *   =  - \beta ^ *  \varepsilon _1 ,\tau _1  =  - \beta \varepsilon _1 ,\delta \tau _{1}^ *   =  - \beta ^ *  \delta _{1} ,and,\delta \tau _{1}  =  - \beta \delta_1.
\end{equation}
Substituting $(A.2)$ into $(A.1)$, Eq. (13) yields with application of $q$-algebra:
\begin{equation}
\begin{array}{l}
 \exp _{q^ *  } \left[ {\tau _1^ *   + \delta \tau _{1}^ *  } \right] = \exp _{q^ *  } \left[ {\tau _1^ *  } \right] \otimes _{q^ *  } \exp _{q^ *  } \left[ {\delta \tau _{1}^ *  } \right] \\
   \\
   \Rightarrow\left\{ {\left( {\exp _{q^ *  } \left[ {\tau _1^ *  } \right]} \right)^{1 - q^ *  }  + \left( {\exp _{q^ *  } \left[ {\delta \tau _{1}^ *  } \right]} \right)^{1 - q^ *  }  - 1} \right\}^{\frac{1}{{1 - q^ *  }}}  \\
   \\
   = Z^ *  \exp \left[ {\tau _1  + \delta \tau _1 } \right] \\

  \\
  \Rightarrow 1 - \left( {1 - q^ *  } \right)\beta ^ *  \left( {\varepsilon _i  + \delta_1 } \right) = [ Z^ *  \exp \left[ {\tau _1  + \delta \tau _1 } \right]]^{1 - q^ *  }  \\
  \\
  \Rightarrow
\beta ^ *   = -\frac{{\ln _{q^ *  } \left\{ {Z^ *  \exp \left[ { - \beta \left( {\varepsilon _1  + \delta _1 } \right)} \right]} \right\}}}{{\varepsilon _1  + \delta _1 }}. \\
 \end{array}
\end{equation}
For consistency of the constant values of $\beta^*$, $\beta$, and $Z^*$, it is required that:
\begin{equation}
\begin{array}{l}
 \beta ^ *   = -\frac{{\ln _{q^ *  } \left\{ {Z^ *  \exp \left[ { - \beta \varepsilon _1 } \right]} \right\}}}{{\varepsilon _1 }} = -\frac{{\ln _{q^ *  } \left\{ {Z^ *  \exp \left[ { - \beta \left( {\varepsilon _1  + \delta _1 } \right)} \right]} \right\}}}{{\varepsilon _1  + \delta _1 }}, \\
 \\
 and \\
 \\
 \beta ^ *   = -\frac{{\ln _{q^ *  } \left\{ {Z^ *  \exp \left[ { - \beta \varepsilon _1 } \right]} \right\}}}{{\varepsilon _1 }} = -\frac{{\ln _{q^ *  } \left\{ {Z^ *  \exp \left[ { - \beta \left( {\varepsilon _1  + \delta _2 } \right)} \right]} \right\}}}{{\varepsilon _1  + \delta _2 }} \\
 \end{array}
\end{equation}

\begin{figure*}[1]
\includegraphics{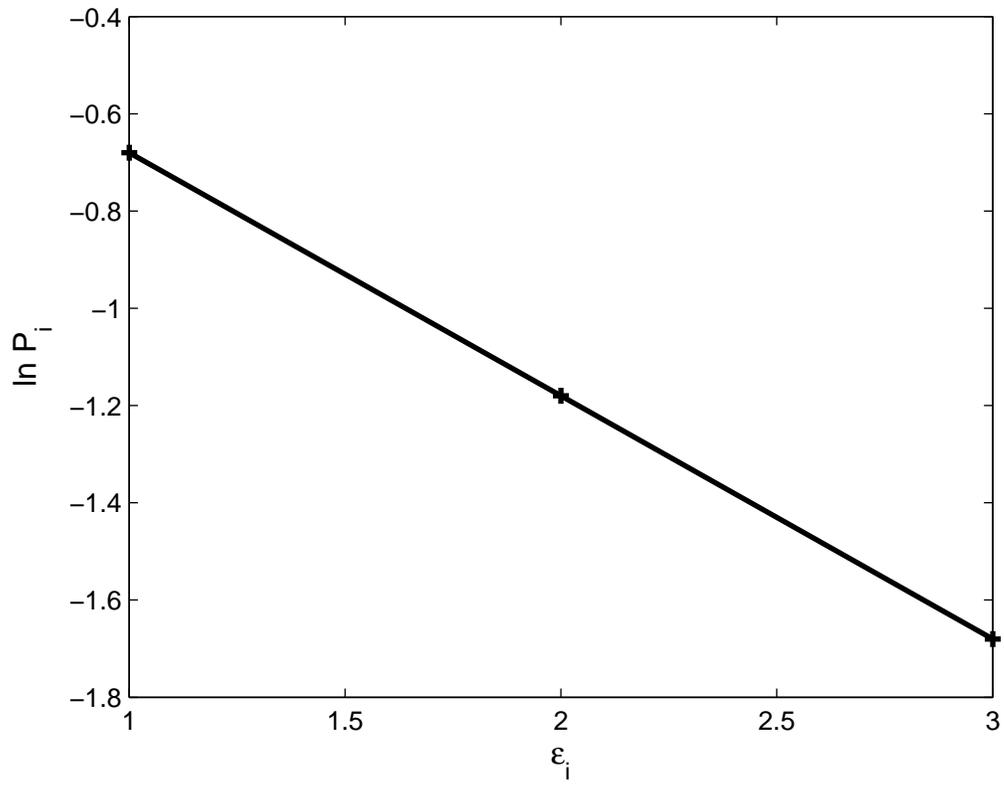}
\caption{\label{fig:epsart} $\varepsilon_i=\{1.0, 2.0, 3.0\},i=1,2,3$ vs. $\ln P_i$ from Eq. (30); $\beta=0.5$.}
\end{figure*}

\begin{figure*}[2]
\includegraphics{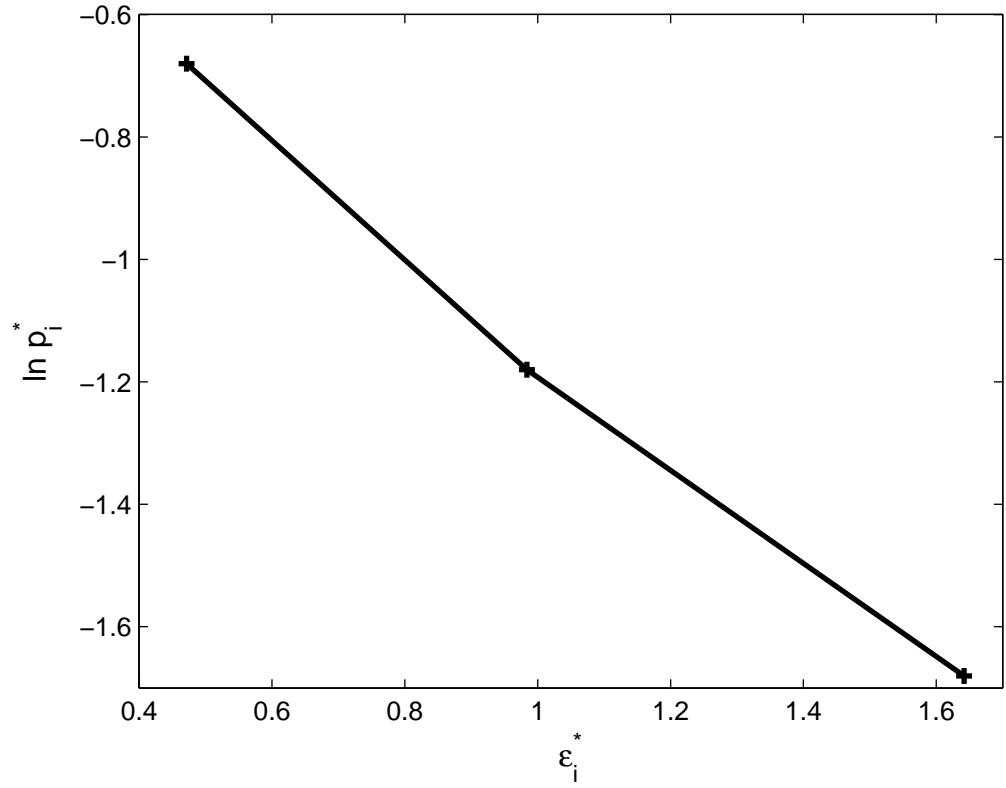}
\caption{\label{fig:epsart} $\varepsilon_i^*=\{0.4713, 0.9839,  1.6420 \}$ vs. $\ln p_i^*$ from Eq. (31); $\beta^*=1.5, q^*=1.5$. }
\end{figure*}

\end{document}